\documentclass[final]{elsart}

\usepackage{graphicx}
\usepackage{amssymb}

\begin{document}

\begin{frontmatter}

\title{
  Quality factor analysis and optimization of 
  digital filtering signal reconstruction 
  for liquid ionization calorimeters
}

\author{Marco Delmastro}

\address{
  European Laboratory for Particle Physics (CERN),
  CH-1211 Geneva 23, Switzerland
}

\ead{Marco.Delmastro@cern.ch}

\begin{abstract}
  The Optimal Filtering (OF) reconstruction of the sampled signals
  from a particle detector such as a liquid ionization calorimeter
  relies on the knowledge of the normalized pulse shapes. This
  knowledge is always imprecise, since there are residual differences
  between the true ionization pulse shapes and the predicted ones,
  whatever the method used to model or fit the particle--induced
  signals.  The systematic error introduced by the residuals on the
  signal amplitude estimate is analyzed, as well as the effect on the
  quality factor provided by the OF reconstruction. An analysis method
  to evaluate the residuals from a sample of signals is developed and
  tested with a simulation tool. The correction obtained is showed to
  preserve the original amplitude normalization, while restoring the
  expected $\chi^2 $--like behavior of the quality factor.
\end{abstract}

\begin{keyword}
  Calorimeters\sep
  Signal Processing\sep
  Digital Filtering
  \PACS 
  29.40.Vj 
\end{keyword}

\journal{{\em Nucl. Instrum. Methods} A}

\end{frontmatter}


\newcommand{\vectornotation}[1]{\ensuremath{\mathbf{#1}}}

\newcommand{\Ns}  {\hbox{$N_{\rm samples}$}}
\newcommand{\NN}  {\hbox{$\mathcal{N}$}}
\newcommand{\vh}  {\hbox{$\vectornotation{h}$}}
\newcommand{\vg}  {\hbox{$\vectornotation{g}$}}
\newcommand{\vr}  {\hbox{$\vectornotation{r}$}}
\newcommand{\va}  {\hbox{$\vectornotation{a}$}}
\newcommand{\vb}  {\hbox{$\vectornotation{b}$}}
\newcommand{\vs}  {\hbox{$\vectornotation{s}$}}
\newcommand{\vn}  {\hbox{$\vectornotation{n}$}}
\newcommand{\vu}  {\hbox{$\vectornotation{u}$}}
\newcommand{\CC}  {\hbox{$\vectornotation{C}$}}
\newcommand{\RR}  {\hbox{$\vectornotation{R}$}}
\newcommand{\II}  {\hbox{$\vectornotation{I}$}}
\newcommand{\vx}  {\hbox{$\vectornotation{x}$}}
\newcommand{\vq}  {\hbox{$\vectornotation{q}$}}
\newcommand{\vy}  {\hbox{$\vectornotation{y}$}}
\newcommand{\PSI} {\hbox{$\vectornotation{\Psi}$}}
\newcommand{\HH}  {\hbox{$\vectornotation{H}$}}
\newcommand{\vcc} {\hbox{$\vectornotation{c}$}}
\newcommand{\vdd} {\hbox{$\vectornotation{d}$}}
\newcommand{\vmm} {\hbox{$\vectornotation{m}$}}
\newcommand{\vkk} {\hbox{$\vectornotation{k}$}}
\newcommand{\RhonN}{\hbox{$\vectornotation{C}(\vn, \NN)$}}
\newcommand{\RhomN}{\hbox{$\vectornotation{C}(\vmm,\NN)$}}
\newcommand{\rhonN}{\hbox{$\vectornotation{\rho}(\vn, \NN)$}}
\newcommand{\rhomN}{\hbox{$\vectornotation{\rho}(\vmm,\NN)$}}


\section{Introduction}
\label{sec:Intro}

The signals arising from the ATLAS electromagnetic calorimeter (EMC)
\cite{LArTDR} are shaped by a bipolar filter, then sampled every 25~ns
at the LHC bunch crossing frequency and stored in analog buffers. Upon
a positive decision from the level-1 trigger, a limited number of
these samples (typically 5) are digitized and acquired. The amplitude
and timing information of the shaped signals are determined combining
the signal pulse samples with a digital filtering technique commonly
called Optimal Filtering (OF) \cite{Cleland94}: this method is
optimized to minimize the noise contribution to the variance of the
reconstructed signal amplitude, while guaranteeing that the latter is
an unbiased estimator of the true amplitude.

Alongside the amplitude and timing information, the OF reconstruction
is designed to produce a quality factor that should allow the
discrimination of pathological signals from regular ones. The
normalized OF quality factor obtained from regular signals follows a
standard $\chi^2$ distribution, while spurious signals generate large
quality factor values: these signals could in principle identified and
be rejected with a cut on this quantity.

The computation of the Optimal Filtering Coefficients (OFC's) for a
given readout cell requires the knowledge of the signal pulse shape
and of the (thermal and pileup) noise time autocorrelation
\cite{Cleland94}. While the latter can be directly measured from
dedicated noise calibration runs and minimum bias events, several
different approaches have been proposed to predict the ATLAS EMC
ionization pulse shapes and their relative amplitudes with respect to
the calibration signals used to probe the detector readout properties
\cite{ATL-LARG-PUB-2001-008,ATL-LARG-PUB-2005-001,JINST-06,ATL-LARG-PUB-2007-010}.
The precision of these pulse prediction methods is quoted in terms of
the difference between the predicted ionization signal and the
observed one, the two pulses being normalized to the same
amplitude. The vector of differences computed for each digitized
sample is commonly called the \emph{residuals} vector (defined in
Equation~\ref{eq:Residuals}): an accurate prediction method is usually
quoted to lead to a difference $<$~1\% at the sample closest to the
signal peak, and always between $\pm$~2\% for the neighboring samples
\cite{ATL-LARG-PUB-2001-008,ATL-LARG-PUB-2005-001,JINST-06,ATL-LARG-PUB-2007-010}.

Assuming that such a precision is achieved for the readout cells of
the ATLAS EMC using a given pulse prediction scheme, this work aims to
study how the unavoidable presence of the residuals systematically
affects the signal amplitude reconstruction and its noise variance
(Section~\ref{sec:OFAmpErr}), as well as the relative quality factor
distribution, thus impairing the discriminating power of the latter
(Section~\ref{sec:Qanalysis}).

A technique to optimize the quality factor without spoiling the
initial reconstructed amplitude normalization is developed and tested
on a reference cell for different possible distributions of the
amplitude, which is proportional to the deposited energy. The same
technique proves to be a powerful tool to extract from data the
ionization pulse shape (up to a normalization factor) when no previous
knowledge -- even approximate -- of it is available
(Section~\ref{sec:PseudoResCorr}).

This technique has been developed in the framework of the ATLAS EMC,
but it holds for any other detector readout system that exploits the
OF reconstruction of multiple--sampled signals (e.g. the ATLAS Tile
hadronic calorimeter \cite{ATL-TILECAL-2005-001}).

\section{Notation and nomenclature}

In the following a matricial notation will be used in all
calculations, $\Ns$ being the number of signal samples digitized
during the data acquisition, thus the typical size of all vectors and
matrices:

\newcommand{\notationlistbegin}{
   \begin{list}{$\bullet$}
     { \setlength{\itemsep}    {1pt}      
       \setlength{\parsep}     {5pt}
       \setlength{\topsep}     {5pt}
       \setlength{\partopsep}  {0pt}
       \setlength{\leftmargin} {3em}
       \setlength{\labelwidth} {3em}
       \setlength{\labelsep}   {0.7em} } 
}

\newcommand{\notationlistend}{
    \end{list}  }

\notationlistbegin

  \item[$\vectornotation{A}$] matrix of size $\Ns\times\Ns$

  \item[$A_{ij}$] ($i$, $j$)-th entry of matrix $\vectornotation{A}$ 

  \item[$\vectornotation{a}$] vector of size $\Ns$

  \item[$a_{i}$] $i$-th entry of vector $\vectornotation{a}$ 

  \item[$\vectornotation{A}^{-1}$] inverse matrix 

  \item[$\vectornotation{A}^{T}$] transposed matrix 

  \item[$\vectornotation{a}^{T}$] transposed vector 

  \item[$a$] scalar 

\notationlistend

\noindent The following nomenclature is used:

\notationlistbegin

 \item[$\vh$] vector of samples of normalized observed ionization signal 

 \item[$\vg$] vector of samples of normalized predicted ionization signal   

 \item[$\vg'$] vector of samples of normalized predicted ionization signal derivative   

 \item[ $\va$] vector of amplitude OFC's computed from $\vg$ (and
   $\vg'$, if time constraint is included \cite{Cleland94})

 \item[$\vb$] vector of time OFC's computed from $\vg$ and $\vg'$, if time constraint is included 

 \item[$\vr$] vector of pulse residuals: 
 
   \begin{eqnarray}
     \vr & = & \vh - \vg 
     \label{eq:Residuals}
   \end{eqnarray}

 \item[$\vs$] vector of observed ionization signal samples for a given pulse amplitude $A$:
   
   \begin{eqnarray}
     \vs & = & A \vh + \vn
     \label{eq:Samples}
   \end{eqnarray}

 \item[$\vn$] vector of noise contributions to signal samples, having the properties:

   \begin{eqnarray}
     \langle \vn \rangle & = & \vec{0} \\
     \langle \vn \vn^T \rangle  & = & \CC
   \end{eqnarray}

 \item[$\CC$] noise covariance matrix:

   \begin{eqnarray}
     \CC & = & \CC^T \\ 
     \CC & = & \sigma_n^2 \RR
   \end{eqnarray}

 \item[$\RR$] weight matrix, built from the noise autocorrelation function:

   \begin{eqnarray}
     \left(\sigma_n^2 \RR \right)^{-1}= \frac{1}{\sigma_n^2} \RR^{-1}
   \end{eqnarray}

 \item[$\II$] identity matrix

\notationlistend

\section{Numerical examples}

All the equations derived in this work are illustrated using a
simulation tool that can generate pulses $\vs$ for a given signal
$\vh$, noise autocorrelation $\RR$ and width $\sigma_n$, and a chosen
distribution of amplitudes $A$. The tool computes OFC's $\va$ (and
$\vb$) from a given pulse prediction $\vg$ and noise autocorrelation
$\RR_{\rm OFC}$ (not necessarily equal to the signal noise
autocorrelation), and applies them to the generated samples to obtain
the corresponding distribution of amplitude estimates $\tilde{A}$ (and
time estimates $\tau$), and the relative quality factors (defined
below in Section~\ref{sec:Qdef}).

\begin{figure}[t]
  \begin{center}  
    \includegraphics[width=0.8\columnwidth]
		    {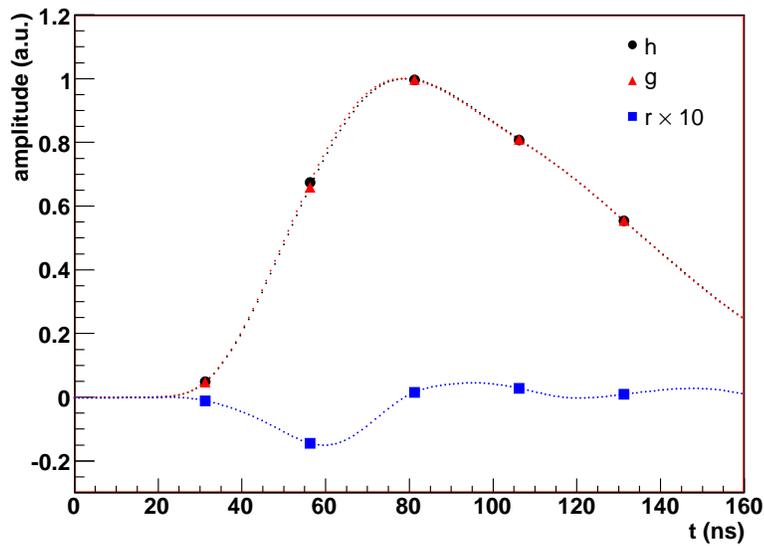}
    \caption{
      Signals  $\vh$ and  $\vg$ and  their  residuals $\vr$ (magnified by
      a factor 10) used in the numerical examples.
    }
    \label{fig:Signals}
  \end{center}
\end{figure}

The test signals $\vh$ and $\vg$ (and their residuals $\vr$) used in
the simulations are plotted in Figure~\ref{fig:Signals}: they
correspond to ionization pulse predictions used during the ATLAS EMC
Barrel commissioning operations in 2007, namely to the ones
corresponding to the middle compartment cells located%
\footnote{The position of ATLAS EMC readout cells is specified by
  using indexes corresponding to the local granularity in
  pseudorapidity $\eta$ and azimuthal angle $\phi$ \cite{LArTDR}. In
  the case of the middle compartment this is
  $\Delta\eta\times\Delta\phi= 0.025\times0.025$, implying:
  \begin{eqnarray}
    \eta & = & 0.025\cdot(\eta_{\rm cell}+0.5) \\
    \phi & = & 0.025\cdot(\phi_{\rm cell}+0.5)
  \end{eqnarray}
}%
at $[\eta_{\rm cell}, \phi_{\rm cell}] = [20, 50]$ ($\vh$) and
$[\eta_{\rm cell}, \phi_{\rm cell}] = [20, 51]$ ($\vg$). They have
been explicitly chosen to be very similar, in order to mimic an
optimal situation in which the pulse prediction largely satisfies the
precision criteria mentioned in Section~\ref{sec:Intro}.

The numerical examples correspond to the fixed pulse phase illustrated
in Figure~\ref{fig:Signals}, that corresponds to the typical EMC data
taking condition at LHC, when $\Ns$ = 5 and the signals are digitized
so that the third sample is located near the pulse maximum
$\pm$~2~ns. The actual values used in the simulations are tabulated in
Table~\ref{tab:PulseValues}.



\begin{figure}[t]
  \begin{center}  
    \includegraphics[width=0.8\columnwidth]
		    {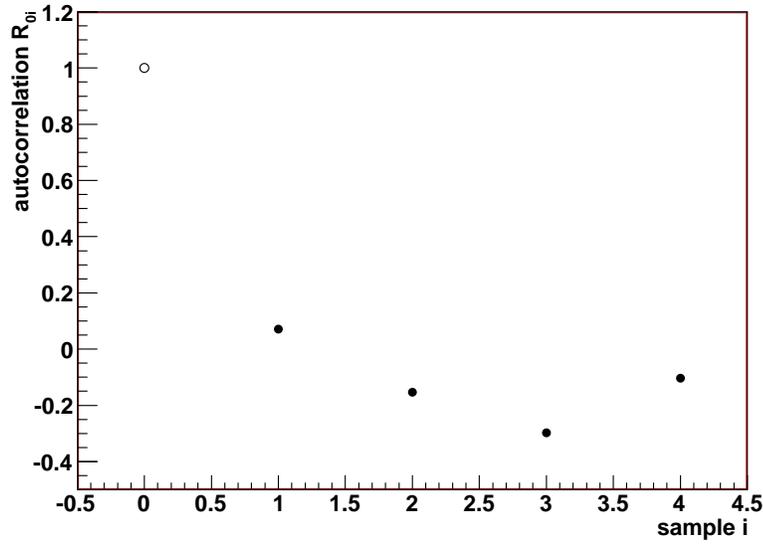}
    \caption{Noise autocorrelation $R_{0i} = R_{j0}$ used in the numerical examples.}
    \label{fig:AutoCorr}
  \end{center}
\end{figure}

Figure~\ref{fig:AutoCorr} shows the values of the thermal noise
autocorrelation function used to generate the noise affecting the
pulses $\vs$. This is carried out by building the matrix $\RR$ as a
Toeplitz matrix based on the relevant autocorrelation function, which
in this case was measured in high gain from middle cell at $[\eta_{\rm
  cell}, \phi_{\rm cell}] = [20, 50]$ during the ATLAS EMC Barrel
commissioning operations in 2007. The matrix $\RR_{\rm OFC}$, used for
computing the OFC's, is based on the measured autocorrelation function
from a neighboring cell at $[\eta_{\rm cell}, \phi_{\rm cell}] = [20,
51]$. The two autocorrelation functions are quite similar, and
previous studies have shown that the residuals obtained are
insensitive to details of the autocorrelation function. We thus use
$\RR_{\rm OFC}$ and $\RR$ interchangeably. A noise width
$\sigma_n$~=~5~ADC counts is used, corresponding to a typical value
for a EMC Barrel middle compartment in high gain.

\begin{figure}[t]
  \begin{center}  
    \includegraphics[width=0.8\columnwidth]
		    {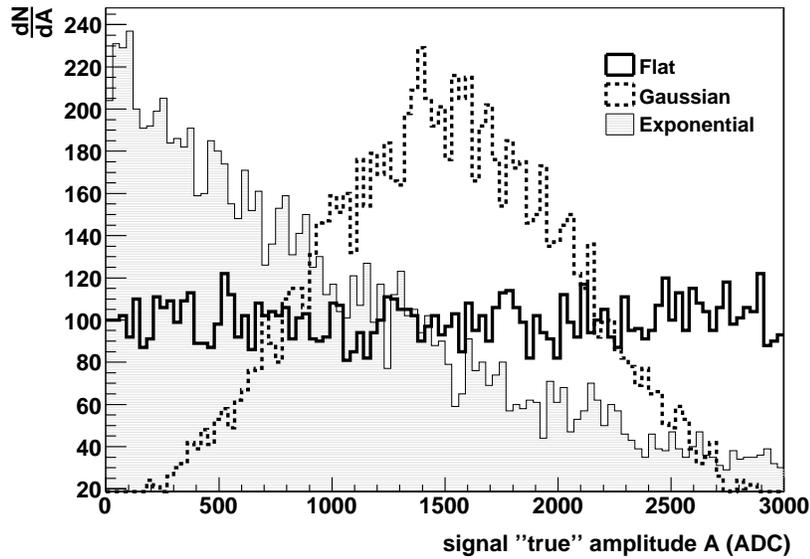}
    \caption{Distributions of the amplitudes $A$ generated for 
      the numerical examples.}
    \label{fig:AmplDist}
  \end{center}
\end{figure}

Figure~\ref{fig:AmplDist} shows the distributions of the amplitudes
$A$ generated for the numerical examples. These amplitudes span the
range between 0 and 3000 ADC counts, covering a realistic range of the
EMC electronics after the ADC pedestal subtraction. In order to
illustrate the effects related to the distribution of the energy
deposited in a EMC cell, we present results from a flat distribution,
a exponential-like and a Gaussian-like distribution of possible
amplitudes $A$.



\section{Amplitude Estimate Error and Noise Variance Bias}
\label{sec:OFAmpErr}

The OF estimate of the signal amplitudes $A$ given the digitized
samples $\vs$ is \cite{Cleland94}:
\begin{equation}
  \tilde{A} = \va^T \vs 
  \label{eq:Amplitude}
\end{equation}
that can be expanded according to Equation~(\ref{eq:Samples}) as:
\begin{equation}  
  \tilde{A} = \va^T \left( A \vh + \vn \right) \\
            = A \va^T \vh + 
	    \NN
\end{equation}
where $\NN = \va^T \vn$ is the (reduced) noise contribution to  OF
amplitude estimate $\tilde{A}$, that has variance:
\begin{equation}
  \sigma_{\NN} = \sigma_n \sqrt{\va^T \RR \va} 
                 \simeq \sigma_n \sqrt{\frac{1}{||\vh||^{2}}}
  \label{eq:OFNoiseApprox}
\end{equation}
where the last approximation holds for small noise autocorrelation and
negligible time jitter, and is exactly true if the noise
autocorrelation is null and the OFC's are not optimized for time
jitter.

\begin{figure}[t]
  \begin{center}  
    \includegraphics[width=0.8\columnwidth]
		    {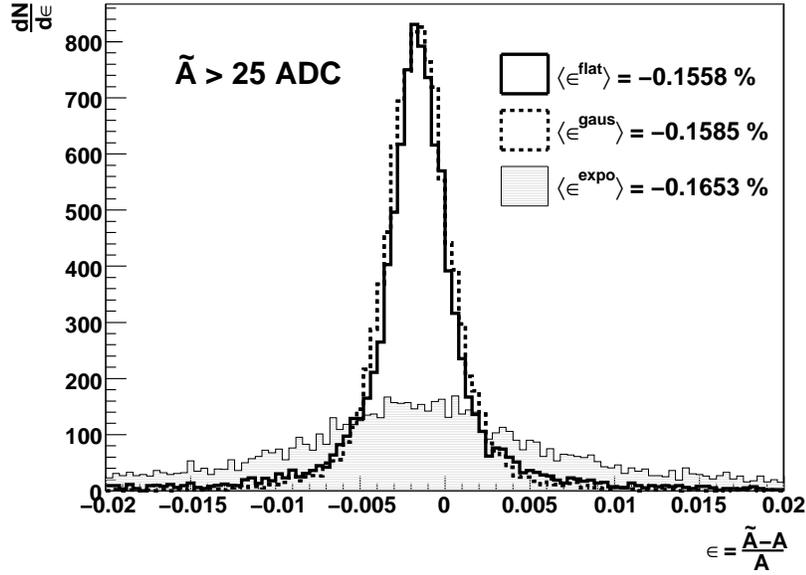}
    \caption{Distributions of the reconstructed amplitude relative
      errors, related to the presence of residuals $\vr$ in the pulse
      used to compute the OFC's.}
    \label{fig:AmplErr}
  \end{center}
\end{figure}

If the  ionization pulse prediction  $\vg$ is not  perfect ($\vg\neq\vh$,
Equation~(\ref{eq:Residuals})), the OF  estimate of the pulse amplitude
suffers of a systematic error $\epsilon=\va^{T}\vr$:
\begin{eqnarray}
  \tilde{A} & = & A \underbrace{\va^T \vg}_{1} + 
	 	  A \underbrace{\va^T \vr}_{\epsilon} 
        	  + \NN_\epsilon \\
  \tilde{A} & = & A \left( 1 + \epsilon \right) + \NN_\epsilon
  \label{eq:OFpeak}
\end{eqnarray}
This error is independent of the pulse amplitude and depends only on
the shape and amplitude of the residuals $\vr$. This property suggests
a simple correction: for a given set of OFC's a scale factor
(e.g. obtained by in--situ calibration with particles of known mass)
would restore the proper scale. In our numerical example
$\epsilon\simeq$~-0.00158: the distributions in
Figure~\ref{fig:AmplErr} confirm that the systematic bias is
independent of the amplitude size and distribution. The average values
of the distributions of the normalized amplitude differences
$\frac{\tilde{A}-A}{A}$ are the same (within statistical errors) for
all the three different forms of the amplitude distribution. The
widths of the $\frac{\tilde{A}-A}{A}$ distributions in
Figure~\ref{fig:AmplErr} differ because of the varying relative impact
of the the noise on the pulse amplitude, since:
\begin{equation}
  \frac{\tilde{A}-A}{A} = \epsilon + \frac{\NN}{A}
\end{equation}
The larger the amount of small signals in the sample under study, the
larger the tails in the $\frac{\tilde{A}-A}{A}$ distribution.

The use of imperfect OFC's also affects the noise contribution to the
OF reconstructed amplitude. The size of this bias can be
approximatively estimated using Equation~(\ref{eq:OFNoiseApprox}), and
recalling that $\va = \frac{\vg}{||\vg||^{2}}$ in the approximation of
negligible noise autocorrelation and when no time constraint is
imposed:
\begin{eqnarray}
  \frac{\sigma_{\NN_\epsilon}}{\sigma_{\NN}} 
  & \simeq & \sqrt{\frac{||\vh||^{2}}{||\vg||^{2}}} 
         =   \sqrt{1+\frac{2\vg^T\vr+||\vr||^{2}}{||\vg||^{2}}} \nonumber \\
  & \simeq &  \sqrt{1+2\epsilon}
    \simeq   1+\epsilon
\end{eqnarray}   
The bias to the minimized variance is of similar magnitude as that
found for the signal amplitude.

\section{Quality Factor Analysis}
\label{sec:Qanalysis}

\subsection{Definition} 
\label{sec:Qdef}

A $\chi^2$-like quantity $Q$ is usually computed from the signal
samples $\vs$ and the amplitude estimate $\tilde{A}$, in order to
evaluate the quality of the OF signal reconstruction. For a signal of
given amplitude $A$ a vector $\vx$ of differences is built as:
\begin{equation}
  \vx = \vs - \tilde{A} \vg
  \label{eq:X}
\end{equation}
and the quality factor is computed as its squared norm:
\begin{equation}
  Q^2 = \vx^T \vx
  \label{eq:Q2}
\end{equation}
or, if the noise autocorrelation is taken into account, as:
\begin{equation}
  Q^2 = \vx^T \RR^{-1} \vx
\end{equation}
If the time constraint is imposed during the OFC's computation, a set
of time OFC's $\vb$ is obtained, and the time estimator $\tau$ can be
computed from the samples as \cite{Cleland94}:
\begin{equation}
 \tau = \frac{1}{\tilde{A}}\vb^T \vs
\end{equation}
In this case, the vector $\vx$ can be computed as:
\begin{equation}
  \vx = \vs - \tilde{A} \left( \vg + \vg' \tau \right )
  \label{eq:Xtime}
\end{equation}

\subsection{Properties and Limitations}
\label{sec:QPropLim}

\begin{figure}[t]
  \begin{center}  
    \includegraphics[width=0.8\columnwidth]
		    {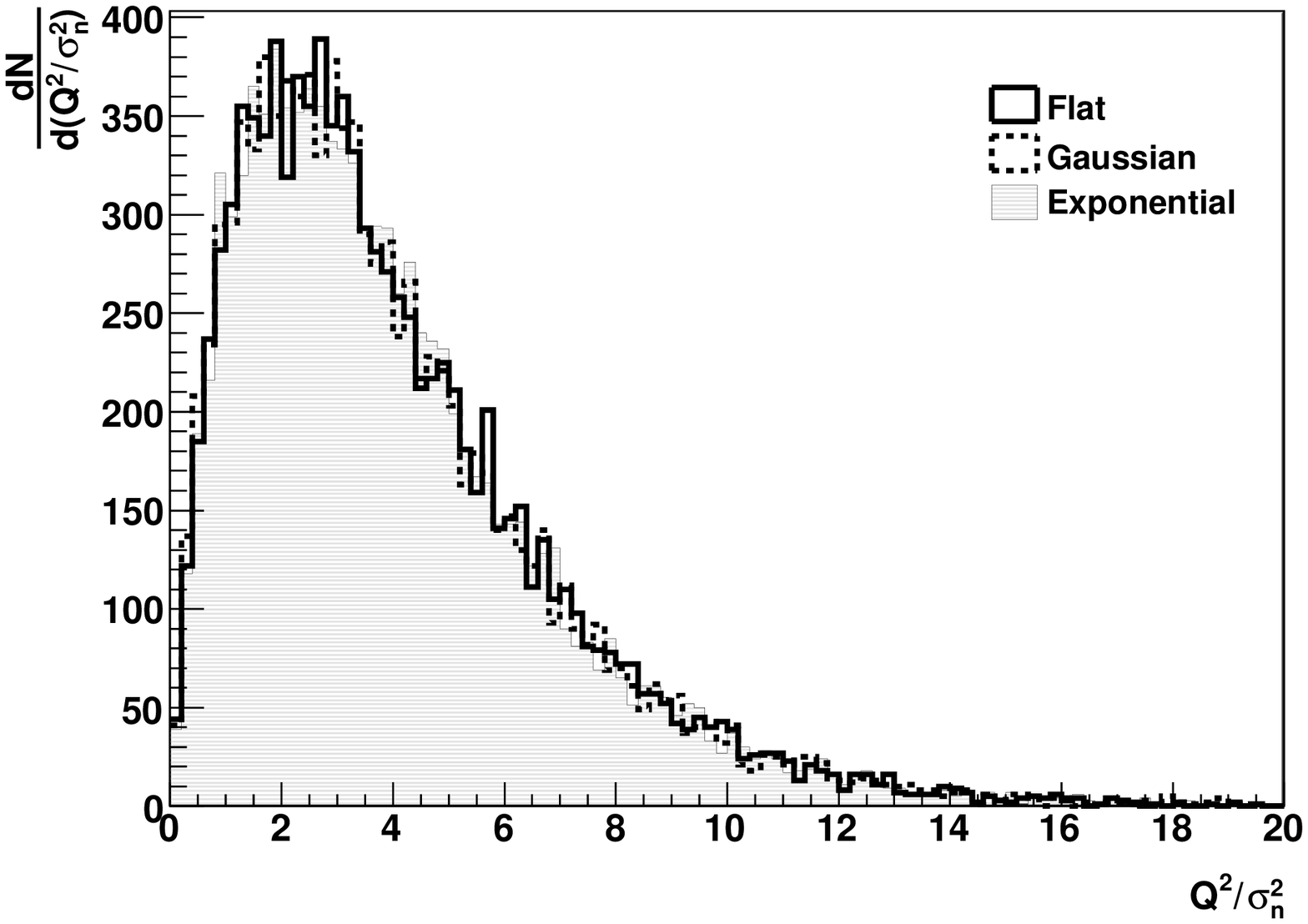}
    \caption{Quality factor distributions for the different signal
      amplitudes, as obtained when using perfect OFC's 
      (i.e. computed from the exact
      ionization pulse $\vh$ for null residuals~$\vr$).}
    \label{fig:Q2DistPerf}
  \end{center}
\end{figure}

The vector of differences $\vx$ as defined in (\ref{eq:X}) can be expanded as:
\begin{eqnarray}
  \vx & = & \vs - \tilde{A} \vg \\
     & = & A \vh + \vn - \va^T \left( A \vh + \vn \right) \vg \nonumber \\
     & = & A \vh + \vn - A \va^T (\vg+\vr) \vg - \va^T \vn \vg \nonumber \\
     & = & A \vh + \vn - A \vg - A \va^T \vr \vg - \NN \vg \nonumber \\
     & = & \vn - \NN \vg + A (\vr - \epsilon \vg) 
     \label{eq:Xexpanded}
\end{eqnarray}
If the ionization pulse prediction is perfect ($\vh=\vg$) the 
difference vector $\vx$ reduces to:
\begin{equation}
   \vx = \vn - \NN \vg
   \label{eq:Xnores}
\end{equation}
and its components $x_i$ are then normally distributed (following the
statistical properties of $\vn$ and $\NN$) around 0. The variable
$Q^2_n$ (\emph{normalized quality factor}):
\begin{equation}
   Q^2_n = \frac{1}{\sigma_n^2} Q^2 =  \frac{1}{\sigma_n^2} \vx^T \RR^{-1} \vx 
   \label{eq:Q2n}
\end{equation}
follows then a $\chi^2$ distribution with $\Ns-1$ degrees of freedom
(Figure~\ref{fig:Q2DistPerf}). The statistical properties of the
normalized $\chi^2$ distribution can then be used to identify, and
optionally reject, distorted signals. For instance in the numerical
examples, where $\Ns~-~1~=~4$, the probability that a signal affected
only by noise gives a normalized quality factor larger than 25 is
smaller than 10$^{-4}$: cutting all signals with $Q^2_n>25$ would
guarantee a very high efficiency while rejecting a large fraction of
spurious signals.



\begin{figure}[t]
  \begin{center}  
    \includegraphics[width=0.8\columnwidth]
		    {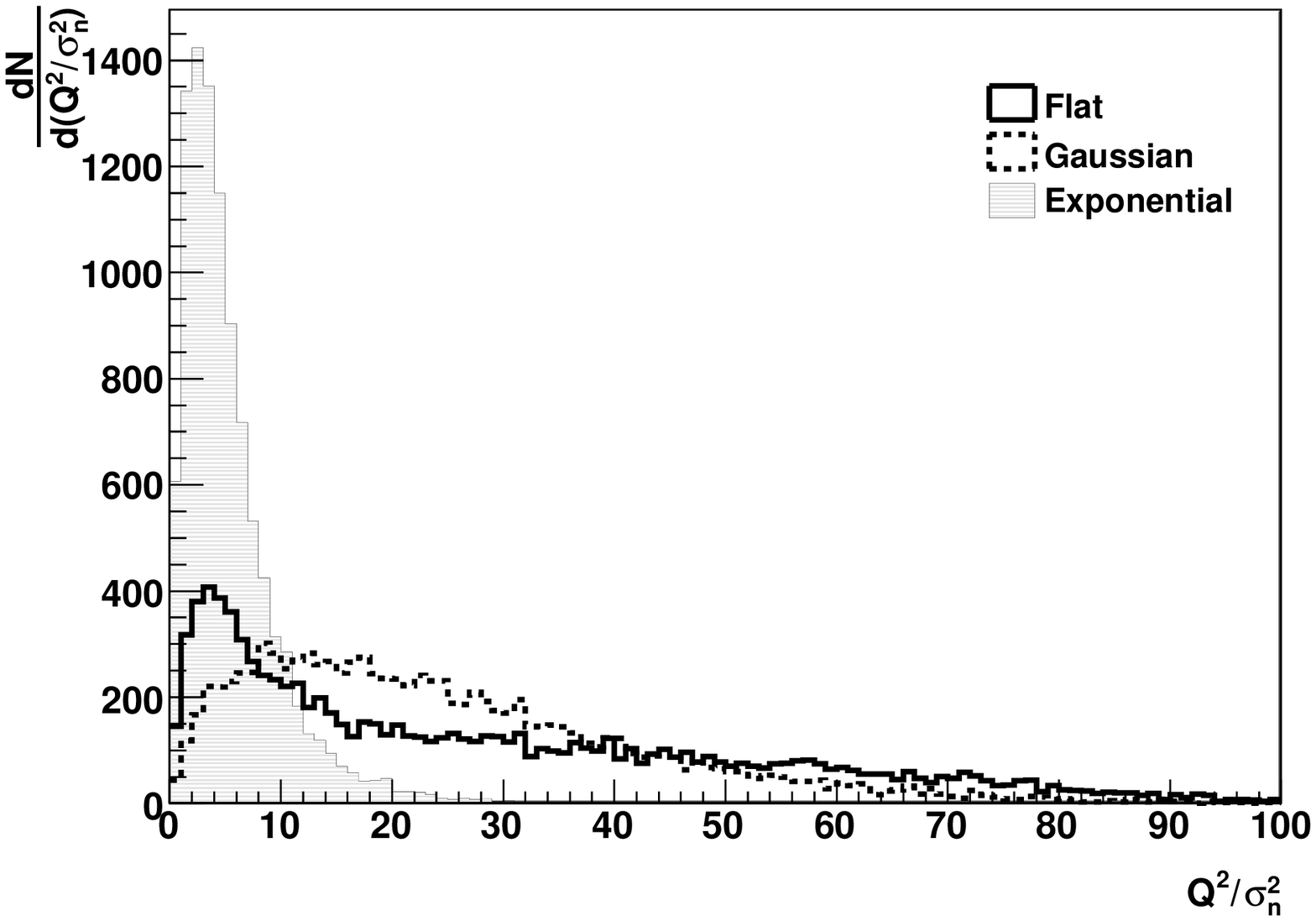}
    \caption{Quality factor distributions for the different signal
      amplitudes, as obtained when using imperfect OFC's 
      (i.e. computed from the predicted  
      ionization pulse $\vg$ for non null residuals $\vr$).}
    \label{fig:Q2DistResi}
  \end{center}
\end{figure}

\begin{figure}[t]
  \begin{center}  
    \includegraphics[width=0.8\columnwidth]
                    {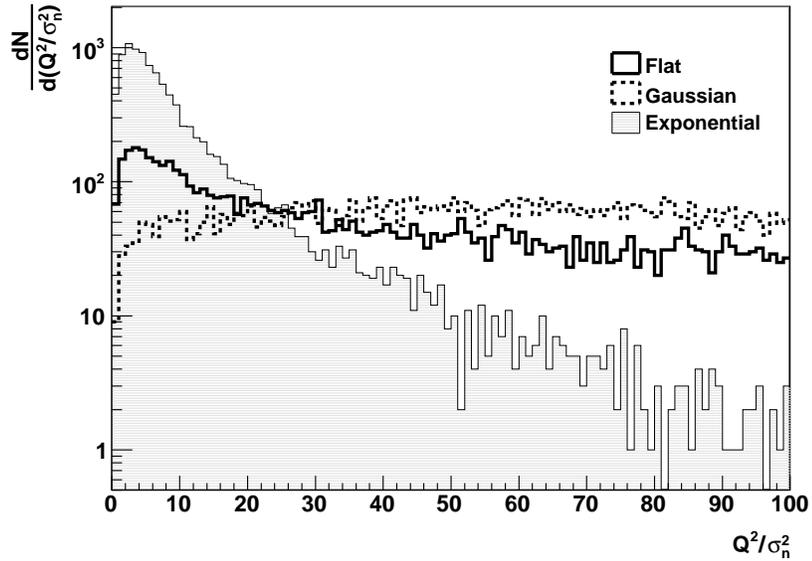}
    \caption{Quality factor distributions obtained using perfect OFC's
      for signals out-of-time by $\sim$~1~ns.}
    \label{fig:QSpuriousMisalign}
  \end{center}
\end{figure}

\begin{figure}[t]
  \begin{center}  
    \includegraphics[width=0.8\columnwidth]
		    {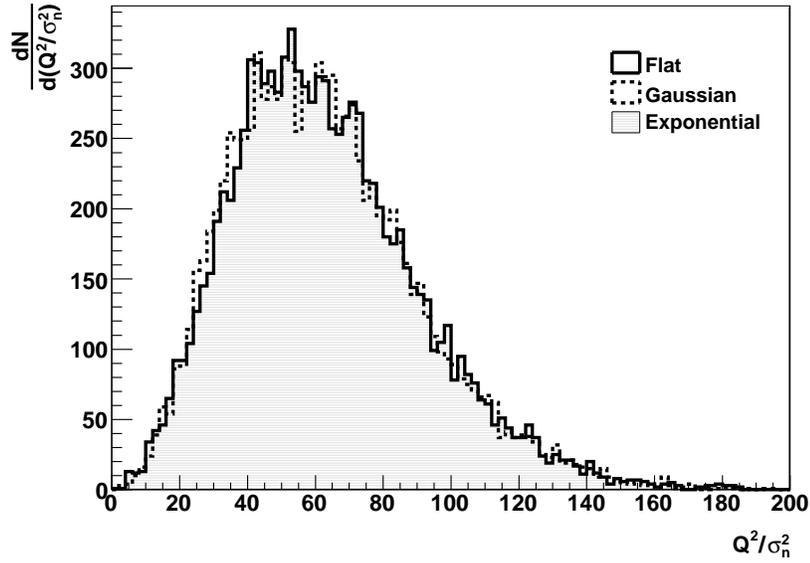}
    \caption{Quality factor distributions obtained using perfect OFC's
      for signals affected by noise bursts. The noise bursts are
      simulated by summing a noise spike -- normally generated with
      a 4$\sigma_n$ mean and 1$\sigma_n$ RMS -- to one signal sample
      at random, and subtracting it from the following sample.}
    \label{fig:QSpuriousNoise}
  \end{center}
\end{figure}

\begin{figure}[t]
  \begin{center}  
    \includegraphics[width=0.8\columnwidth]
		    {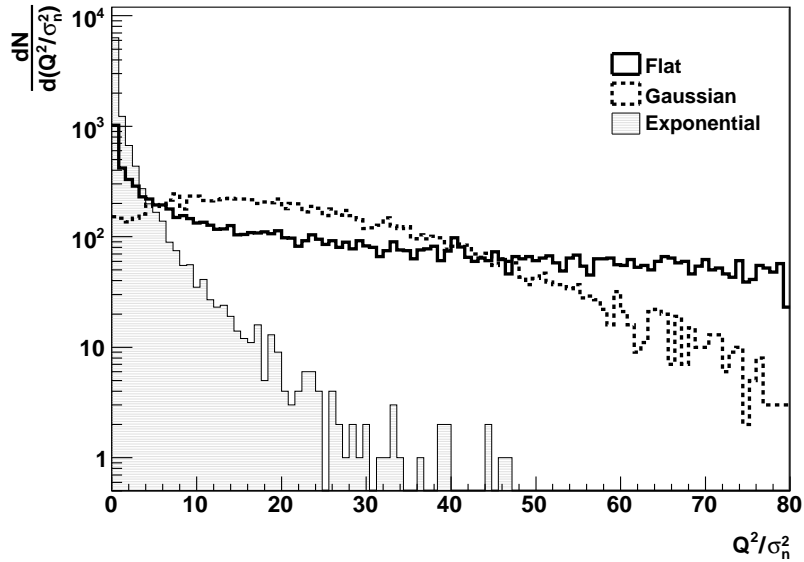}
    \caption{Quality factor distributions obtained using perfect OFC's
      for signals affected by large capacitive cross-talk from
      neighbor cells. The cross-talk distortion is simulated by
      summing to the current pulse with one having the shape of the signal
      derivative $\vh'$, and amplitude proportional to 1.5\% of a
      signal independently generated from the same amplitude
      distribution.}
    \label{fig:QSpuriousXtalk}
  \end{center}
\end{figure}

On the other hand, if the pulse prediction is not perfect
($\vh\neq\vg$) the $\vx$ components are no longer normally
distributed, since they all contain a contribution related to the
distribution of the pulse amplitude $A$
(Equation~(\ref{eq:Xexpanded})). In this case the $Q^2_n$ variable
defined in Equation~(\ref{eq:Q2n}) no longer follows a normalized
$\chi^2$ distribution, long tails arise in the distributions
(Figure~\ref{fig:Q2DistResi}), and the shape of the normalized quality
factor $ Q^2_n$ depends now on the original signal amplitude
distribution. In this case the discriminating power of the quality
factor is impaired. If in fact the normalized quality factor of
strongly distorted signals (e.g. in the rare case of important signal
pileup from previous events) is so large ($\gtrsim~10^{3}$) that even
the use of imperfect OFC's would not really affect its rejection
capability, there are other categories of spurious signals for which
the distortion is smaller, that produce quality factor values
comparable to the ones obtained using imperfect OFC's. This is for
instance the case of slightly out-of-time signals
(Figure~\ref{fig:QSpuriousMisalign}); of signals affected by noise
bursts, when occasionally a large noise spike in the readout
electronics coherently distorts the signal positively in one of the
samples and negatively in the following one
(Figure~\ref{fig:QSpuriousNoise}); of large cross--talk from
neighboring cells (Figure~\ref{fig:QSpuriousXtalk}), when a signal
similar to the pulse derivative (capacitive cross--talk), and
proportional to the energy release around the cell, distorts the pulse
\cite{ATL-LARG-2000-007}. In these conditions, when using perfect
OFC's the large quality factor values produced are of a similar
magnitude as the ones obtained from regular signals using imperfect
OFC's.





Another problem concerns the variability of the quality factor
distributions from cell to cell. The residuals may in fact differ from
cell to cell due to differences in the properties of the readout
channels, as well as the accuracy of the pulse predictions and OFC
calculations for each cell. The distribution of amplitudes also
differs from cell to cell, due to the spatial variation of energy
deposits, and of the different size and depth of the cells in the
detector. Differences in the quality factor distributions complicate
cross-comparisons of this quantity for different cells, or the
construction of meaningful cumulative distributions.

\section{Residuals estimation from quality factor components}
\label{sec:PseudoResCorr} 

\subsection{Pseudo-residuals definition}

For a pulse of given amplitude $A$ a vector $\vy$ can be built as the
ratio of the $\vx$ components (\ref{eq:X}) and the reconstructed
amplitude $\tilde{A}$ (\ref{eq:OFpeak}):
\begin{equation}
  \vy = \frac{\vx}{\tilde{A}} 
      = \frac{\vn-\NN\vg+A(\vr-\epsilon\vg)}{A(1+\epsilon)+\NN}
  \label{eq:PseudoResNoise}
\end{equation}
We call \emph{event pseudo-residuals} the component of the $\vy$
vector computed for a given pulse amplitude and noise contribution
(i.e., for a given energy deposit and readout condition in the cell,
thus for a given event).

If the noise contribution could be ignored for each event
($\sigma_n=0$, $\vn=\vec{0}$, $\NN=0$), this $\vy$ vector would
assume a simplified form, independent of the pulse amplitude:
\begin{equation}
  \PSI = \frac{\vr-\epsilon\vg}{1+\epsilon}
  \label{eq:PseudoRes}
\end{equation}
We call \emph{average pseudo-residuals} this $\PSI$ vector, and
discuss below (Section~\ref{sec:GaussRatio}) how and under which
conditions the components of $\PSI$ can be obtained from the
distributions of the components of $\vy$. Note that the average
pseudo-residuals are orthogonal to the amplitude OFC's, since:
\begin{equation}
  \va^T \PSI = 0
  \label{eq:PseudoResOrtho}
\end{equation}

\subsection{Estimation of residuals and calculation of OFC correction from average pseudo-residuals}

The system of linear equations which follows%
\footnote
{
Equation (\ref{eq:PseudoResLinSystem}) exploits the relation:
\begin{equation}
  \underbrace{
    \left( \vectornotation{a}^T \vectornotation{b} \right)
  }_{\alpha\in\Re}
  \vectornotation{c} =
  \underbrace{
    \left( \vectornotation{c} \vectornotation{a}^T \right)
  }_{\vectornotation{M}\in\Re^n\times\Re^n}
  \vectornotation{b}
\end{equation}
}
from Equation~(\ref{eq:PseudoRes}):
\begin{eqnarray}
  (1 + \epsilon) \PSI & = & \vr - \epsilon \vg \\
  \PSI + \PSI \va^T \vr & = & \vr - \vg \va^T \vr \label{eq:PseudoResLinSystemPre} \\
  \underbrace{
  \left( \II - \vg \va^T - \PSI \va^T \right) 
  }_{\HH}
   \vr & = & \PSI \label{eq:PseudoResLinSystem}
\end{eqnarray}
unfortunately cannot be solved to find $\vr$. In fact the matrix $\HH$
is in general ill--conditioned, and in particular exactly singular,
when no noise autocorrelation is present in the system. This property
is directly related to the component $\vg \va^T$ reducing the rank of
$\HH$, since:
\begin{equation}
  \det(\vg \va^T) \simeq 0
\end{equation}
for general values of $\vr$, and exactly null when $\RR=\II$.

The impossibility to solve system (\ref{eq:PseudoResLinSystem}) is
directly related to the (hopeless) attempt to extract information
about $\vr$ from $\epsilon$ in
Equation~(\ref{eq:PseudoResLinSystemPre}): in fact, given a scalar
$\epsilon$ and a vector $\va$, there unfortunately exists a infinite
set of vectors $\vr$ that satisfy the relation $\va^T\vr=\epsilon$.%
\footnote{
Consider for instance the vector:
\[
\vectornotation{\rho} = \left\{ \frac{\epsilon}{a_0} , 0, 0, ..., 0 \right\} 
\]
} This is an intrinsic limitation of the OF signal reconstruction that
cannot be circumvented: by applying the OFC's $\va$ to the samples
$\vs$ when estimating the signal amplitude a part of information about
the original pulse $\vh$ is lost.
 
On the other hand, from Equation~(\ref{eq:PseudoRes}) it is possible
to define a more general class of the vectors $\vr_\lambda$ that
satisfy the relation $\va^T \vr_\lambda = \lambda$:
\begin{equation}
  \vr_\lambda = (1+\lambda)\PSI +\lambda\vg \qquad \lambda \in \Re
  \label{eq:ResLambda}
\end{equation}
When any of these $\vr_\lambda$ vectors (\ref{eq:ResLambda}) is used
to correct the ionization pulse prediction $\vg$, the resulting
corrected pulse $\vg_c$ is \emph{exactly proportional} to the
ionization pulse $\vh$ we were aiming for, apart from a scale factor:
\begin{equation}
  \vg_c^\lambda = \vg + \vr_\lambda =
  (1+\lambda)(\vg+\PSI) = \frac{1+\lambda}{1+\epsilon} \vh
  \label{eq:PseudoResCorr}
\end{equation}

Of course the preferred choice would be $\lambda=\epsilon$, but the
impossibility of this option has already been discussed. On the other
hand, by choosing $\lambda=0$ an interesting property of the
correction is found: the OFC $\va_c$ obtained from the corrected pulse
$\vg_c^{\lambda=0}$ \emph{preserve the reconstructed signal amplitude
  normalization} obtained from the original $\va$ OFC's obtained from
the initial pulse prediction $\vg$. If we let $\va_h$ represent the
OFC's corresponding to the true ionization pulse $\vh$, we have the
following relationship between the corrected and the ``true'' OFC's:
\begin{equation}
  \va_c = \left( 1+\epsilon \right) \va_h
\end{equation}
and then ($\NN_h = \va_h n$):
\begin{eqnarray}
  \tilde{A}_c & = & \va_c s \\
              & = & \left( 1+\epsilon \right) \left(A \va_h \vh + \va_h \vn\right) \\
	      & = & \left( 1+\epsilon \right) A + \left( 1+\epsilon
              \right)\NN_h \label{eq:PseudoResRecoAmpl}
\end{eqnarray}
The $\va_c$ OFC's \emph{correctly compensate for errors in the initial
  prediction of the ionization pulse shape}, while preserving the
amplitude normalization obtained from the original $\va$ OFC's. When
trying to optimize the global cell calibration this certainly
represents an advantage. In fact, in order to correct for the effects
of material in front of the detector, or to compute the correct global
energy scale, the reconstructed signal amplitudes are always
multiplied by additional calibration factors obtained \emph{in--situ}
from reference data samples (e.g. electrons from $W^{\pm}$ and $Z^{0}$
decays). In the process of computing these factors with an increasing
data sample, and optimizing at the same time the reconstruction of the
signal amplitudes, the normalization of the latter is better kept
preserved in order not to mix the effects.

Whatever the choice of $\lambda$ (and thus also for $\lambda=0$) the
use of the corrected OFC's \emph{restores the proper behavior of the
quality factor}. In fact the components of the difference vector
$\vx_c$:
\begin{eqnarray}
  \vx_c & = & \vs - \tilde{A}_c \vg_c \\
  & = & A \vh + \vn - \frac{1+\epsilon}{1+\lambda} \va_h^T \left( A
  \vh + \vn \right) \frac{1+\lambda}{1+\epsilon} \vh \nonumber \\
  & = & \vn - \NN_h
  \label{eq:Xc}
\end{eqnarray}
are now correctly normally distributed around 0, and:
\begin{equation}
  \frac{1}{\sigma_n^2} Q_c^2 = \frac{1}{\sigma_n^2} \vx_c^T \RR^{-1} \vx_c
\end{equation}
is guaranteed to follow a $\chi^2$ distribution with $\Ns-1$ degrees
of freedom.

\subsection{From event pseudo-residuals to average pseudo-residuals}
\label{sec:GaussRatio}

The possibility of correcting the OFC's by exploiting
Equation~(\ref{eq:PseudoResCorr}) is related to the capability of
extracting consistent estimators of the average pseudo-residuals
$\PSI$ from the distribution of the event pseudo-residuals $\vy$ as
defined in Equation~(\ref{eq:PseudoResNoise}).

For a given value of the signal amplitude $A$ every $y_i$ is
distributed as the ratio of two Gaussian variables with different
non-zero means, different variances, and a certain correlation. The
$\vy$ vector can in fact be expressed as:
\begin{equation}
  \vy(A) = \frac{A \vcc + \vmm }{A d + \NN}
      = \frac{\vcc + \frac{1}{A}\vmm }{d + \frac{1}{A}\NN }
\end{equation}
where:
\begin{equation}
  \begin{array}{rcl}
    \vcc & = & \vr-\epsilon\vg \\
       d & = & 1+\epsilon 
  \end{array}
\end{equation}
and
\begin{equation}
  \begin{array}{rcl}
    \vmm & = & \vn-\NN\vg \\
    \label{eq:PseudoResNominatorNoise}
  \end{array}
\end{equation}
The components of the $\vmm$ vector are normally distributed around 0
with standard deviation $\sigma_{\vmm}$, $\rhomN$ being the vector of
their correlation with the denominator noise factor $\NN$.

The distribution of such a ratio variable is not necessarily Gaussian,
especially in case of noisy measurements. On the other hand under
certain assumptions, namely that the denominator of the ratio is very
unlikely to become negative:
\begin{equation}
  A d \geq 3 \sigma_{\NN}
  \label{eq:PositiveDenominator}
\end{equation}
the probability distribution of this ratio is exactly known%
\footnote{Additionally, under the hypothesis
(\ref{eq:PositiveDenominator}) a transformation exists to compute a
derived variable that is exactly Gaussian.}
\cite{Geary1930,Hinkley1969}, and the distribution of each component
of $\vy(A)$ for a given value of $A$ approaches a Gaussian $G(\langle
y_i(A) \rangle, \sigma_{y_i(A)} )$ with:
\begin{equation}
  \langle \vy(A) \rangle \sim \frac{\vcc}{d} = \PSI
  \label{eq:RatioGaussianMean}
\end{equation}
\begin{equation}
  \sigma_{\vy(A)} \sim 
  \frac{1}{Ad} \sqrt{ \sigma_{\vmm}^2 + \PSI^2 \sigma_{\NN}^2
    - 2 \rhomN \sigma_{\vmm} \sigma_{\NN} }
  \label{eq:RatioGaussianRMS}
\end{equation}
Note that the standard deviation $\sigma_{\vy(A)}$ scales as
$\frac{1}{A}$. 
The local%
\footnote{The \emph{local} attribution refers here to the fixed value
of the pulse amplitude $A$.} normality condition
(\ref{eq:PositiveDenominator}) for $\vy(A)$ is amply satisfied by:
\begin{equation}
   \tilde{A} \ge 3 \sigma_n 
   \label{eq:GaussianityReq}
\end{equation}
since $\sigma_n > \sigma_{\NN}$ because of the OF noise
reduction.

Our specific case is additionally complicated by the fact the
amplitude $A$ in (\ref{eq:PseudoResNoise}) follows its particular
distribution $P(A)$ that is not a priori known and that might change
from cell to cell depending on the position in the detector and the
event sample selection. The cumulative distributions $D(y_i|P(A))$ of
the event pseudo-residual components $y_i$, given a probability
distribution function $P(A)$ for the signal amplitudes and holding the
normality requirements (\ref{eq:GaussianityReq}), are then:
\begin{equation}
  D(y_i|P(A)) = \int dA P(A) G( \psi_i, \sigma_{y_i(A)} )
\end{equation}
While the actual shapes of these $D(y_i|P(A))$ distributions vary
according to $P(A)$ (and in most cases are definitively non-Gaussian!),
it is always true that every $D(y_i|P(A))$ is symmetric, and that the
vector of their means corresponds to the average pseudo-residual
vector whatever the initial $P(A)$, since:
\begin{eqnarray}
  \langle D(y_i|P(A)) \rangle 
  & = & \int dy_i y_i \int dA P(A) G(\psi_i,\sigma_{y_i(A)})\nonumber\\
  & = & \underbrace{\int dA P(A)}_{\equiv 1} 
        \underbrace{\int dy_i y_i G(\Psi_i,\sigma_{y_i(A)}) }_{\equiv
        \Psi_i \, \forall A \, , \, \, {\mathrm{Equation}~(\ref{eq:RatioGaussianMean})} } \nonumber \\
  & = & \Psi_i \label{eq:PseudoResEstimator}
\end{eqnarray}
It is then possible to estimate the average pseudo-residuals $\PSI$ as
the means of the distributions of the event pseudo-residuals, after
having applied the local normality requirements
(\ref{eq:GaussianityReq}) on the events entering the distributions.

\begin{figure}[t]
  \begin{center}  
    \includegraphics[width=0.8\columnwidth]
		    {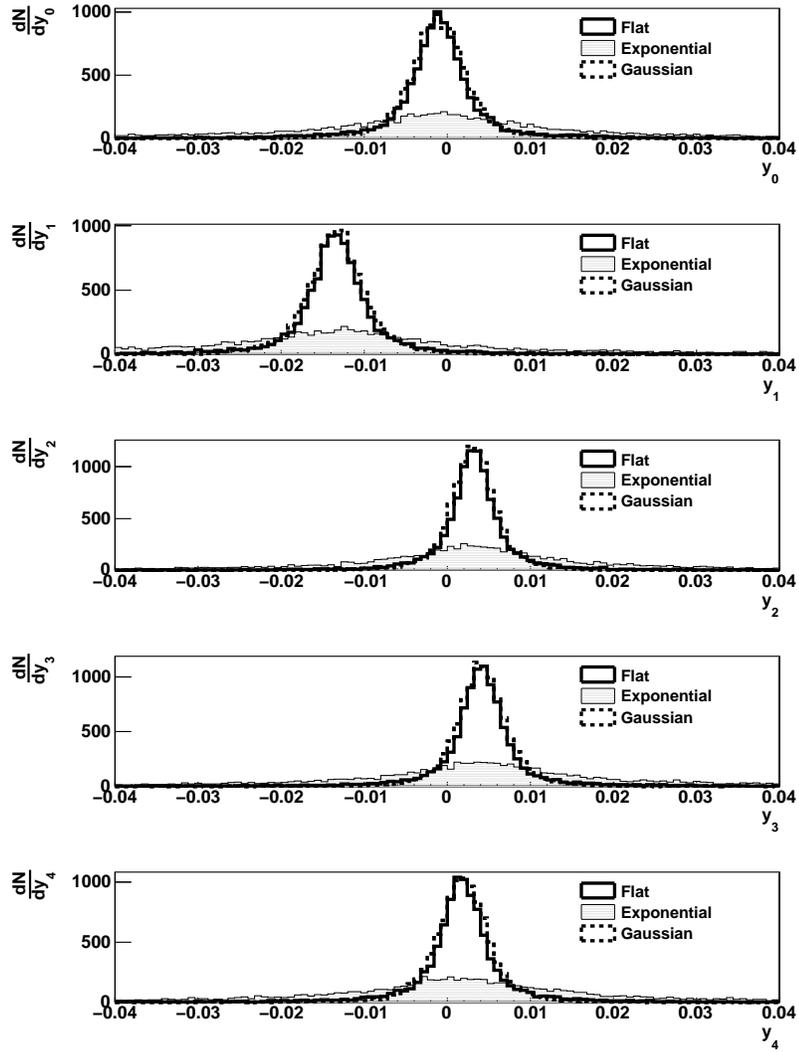}
    \caption{Event pseudo-residual component distributions, as
             obtained for the different signal amplitudes for
             $\tilde{A}>25=5\sigma_n$.}
    \label{fig:PseudoResDists}
  \end{center}
\end{figure}

\begin{figure}[t]
    \begin{center}  
      \includegraphics[width=0.8\columnwidth]
		      {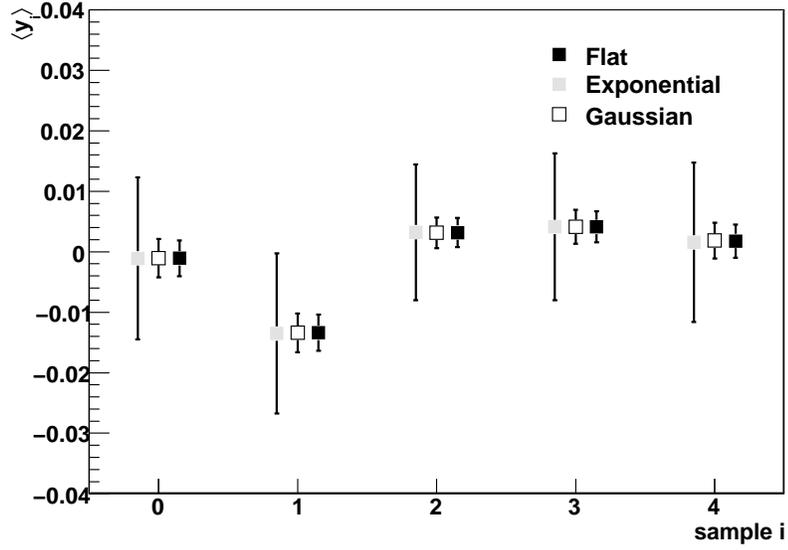}
      \caption{Means of the event pseudo-residual component
               distributions for the different signal amplitude
               distribution $P(A)$ and for $\tilde{A}>25=5\sigma_n$. 
               Error bars indicate the widths of the
               distributions.}
      \label{fig:PseudoResMeans}
    \end{center}
\end{figure}

The property expressed in Equation~(\ref{eq:PseudoResEstimator}) is
clearly verified in our numerical
examples. Figure~\ref{fig:PseudoResDists} shows the event
pseudo-residual distributions corresponding to the three different
functional forms for $P(A)$, and built from events passing the tight
threshold cut on the reconstructed amplitude
$\tilde{A}>25=5\sigma_n$. Their shape certainly depends on the initial
distribution $P(A)$ of the signal amplitudes $A$, but as expected they
are always symmetric, and their means (here evaluated by means of a
Gaussian fit to the distributions) correspond in all cases to the
average pseudo-residual vector $\PSI$ as shown in
Figure~\ref{fig:PseudoResMeans}, the error bars accounting for the
width of the corresponding distribution. It is evident that a
distribution for the ratio $y_i$ corresponding to an amplitude
distribution $P(A)$ with predominately small values of $A$ (such as
the exponential in this example) would have larger tails, as the
prevailing contributors to $D$ are Gaussian distributions of large
standard deviation (see Equation~(\ref{eq:RatioGaussianRMS})).



\subsection{Average pseudo-residual estimators as OFC pulse correction}
\label{sec:OFCCorr}

The mean values defined in Equation~(\ref{eq:PseudoResEstimator}) can
be used as estimator of $\PSI$ to correct the original OFC's via
Equation~(\ref{eq:PseudoResCorr}), by choosing $\lambda=0$ to follow
the prescription to preserve the original reconstructed signal
amplitudes, as discussed in Section~\ref{sec:PseudoResCorr}. Because
of the small size of this correction the variation of the pulse
derivative $\vg'$ can safely be neglected, and the original one can be
used to computed OFC's optimized for time jitter.

\begin{figure}[t]
  \begin{center}  
    \includegraphics[width=0.8\columnwidth]
		    {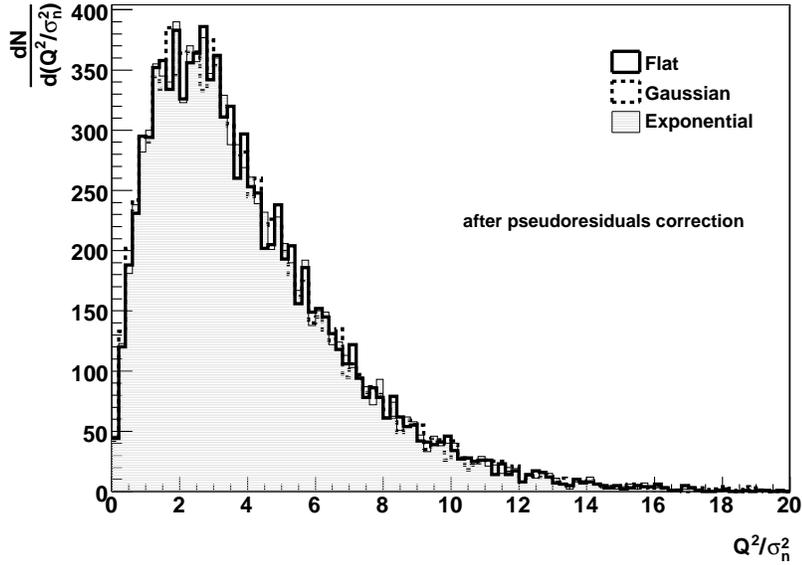}
    \caption{Distributions of reconstructed amplitude relative error,
      as obtained when correcting the OFC's input
      pulse $\vg$ with the pseudo-residual estimators $\langle\vy\rangle$.}
    \label{fig:AmplErrCorr}
  \end{center}
\end{figure}

\begin{figure}[t]
  \begin{center}  
    \includegraphics[width=0.8\columnwidth]
		    {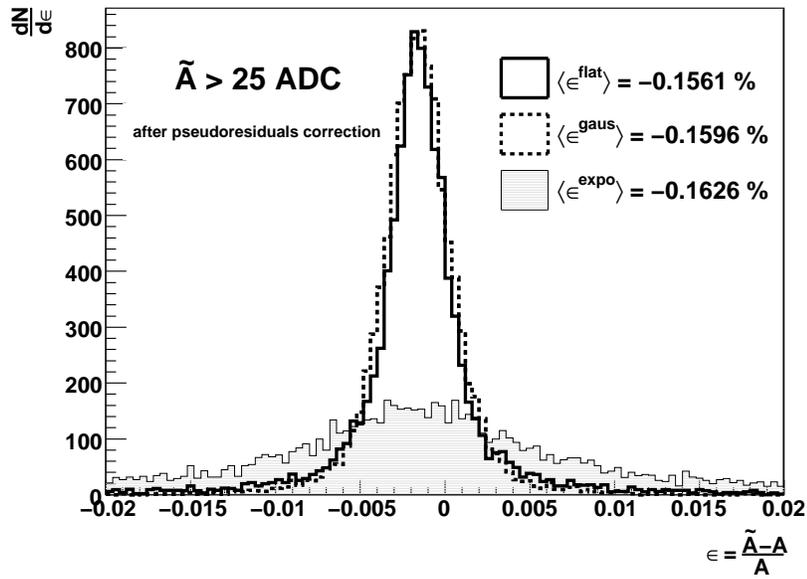}
    \caption{Quality factor distributions for the different signal
      amplitudes, as obtained when correcting the OFC's input
      pulse $\vg$ with the pseudo-residual estimators $\langle\vy\rangle$.}
    \label{fig:Q2DistCorr}
  \end{center}
\end{figure}

When new OFC's are computed from $\vg_c$, $\vg'$ and $\RR$, and
reapplied to the original signal samples $\vs$, as expected from
Equation~(\ref{eq:PseudoResRecoAmpl}) the reconstructed amplitudes are
unchanged, preserving the original normalization
(Figure~\ref{fig:AmplErrCorr}). The corrected quality factors follow
now a $\chi^2$--like distribution, for any initial pulse
amplitude distribution (Figure~\ref{fig:Q2DistCorr}).



\subsection{Pulse shape estimate with flat filter average pseudo-residuals}
\label{sec:FlatFilter}

The pseudo-residuals correction technique proves also to be a powerful
tool to extract from data the ionization pulse shape (up to a
normalization factor) when no previous knowledge of it is available,
whatever the available signal sample. This technique addresses the
relative normalization of the signal samples corresponding to
different energy deposits in the readout channel under study.

\begin{figure}[t]
  \begin{center}  
    \includegraphics[width=0.8\columnwidth]
		    {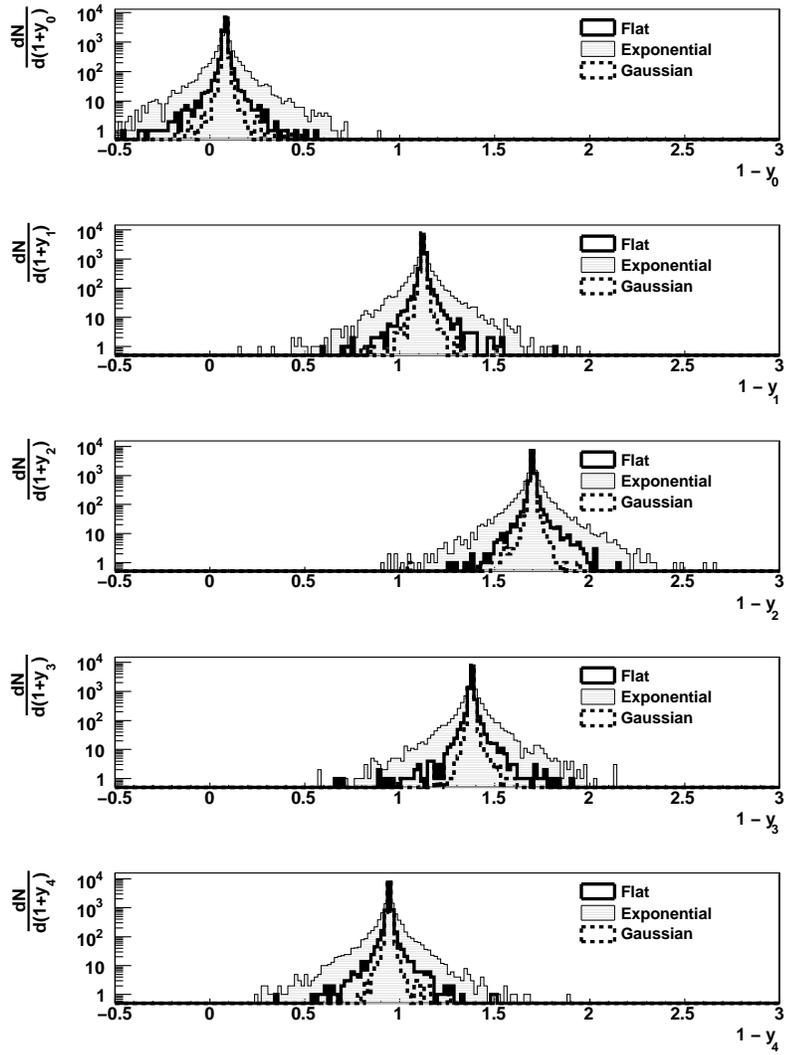}
    \caption{Distributions of event pseudo-residual components biased
      by 1, as obtained for the different signal amplitudes for
      $\tilde{A}>25=5\sigma_n$, when a flat filter is used for the
      amplitude reconstruction.}
    \label{fig:FlatFilterDists}
  \end{center}
\end{figure}

\begin{figure}[t]
    \begin{center}  
      \includegraphics[width=0.8\columnwidth]
		      {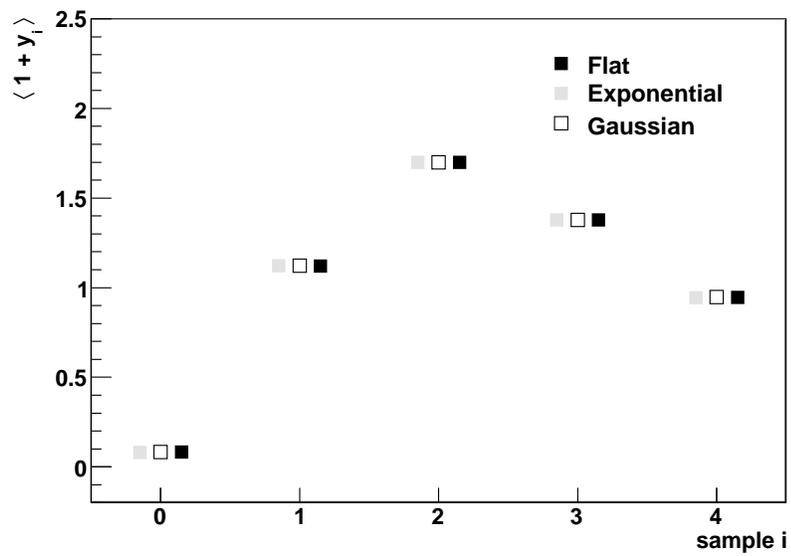}
      \caption{Means of the distributions of the event pseudo-residual
        components biased by 1 for the different signal amplitude
        distribution $P(A)$ and for $\tilde{A}>25=5\sigma_n$, when a
        flat filter is used for the amplitude reconstruction. The
        error bars are smaller than the point markers.}
      \label{fig:FlatFilterMeans}
    \end{center}
\end{figure}

Let $\vu$ be the normalized vector of size $\Ns$ with all components
of equal magnitude (\emph{flat pulse}):
\begin{equation}
  \vu = \left\{ 1, 1, \dots, 1 \right\}
\end{equation}
and compute the corresponding $\va_{u}$ OFC's
(\emph{flat~filter}). When they are used to reconstruct the signal
amplitude and obtain the pseudo-residuals $\PSI_{u}$, from
Equation~(\ref{eq:PseudoResCorr}) one gets:
\begin{equation}
  \vu + \PSI_{u} = \frac{1}{1+\epsilon_{u}} \vh
\end{equation}
that is exactly the ionization pulse $\vh$, up to the normalization
factor $\frac{1}{1+\epsilon_{u}}$ (which, for the reason discussed
above, is again bound to remain unknown).
Figure~\ref{fig:FlatFilterDists} shows the distributions of the event
pseudo-residuals (biased by 1) when the flat filter is used for the
reconstruction of the signals belonging to the example distributions;
the vectors of their averages nicely correspond to the original
ionization pulse $\vh$ (Figure~\ref{fig:FlatFilterMeans}). In our
example $\epsilon_{u}$~=~-0.412544: if this value is used to scale the
biased pseudo-residual averages in order to compare them with the true
$\vh$, excellent agreement, ranging from $\sim10^{-4}$ (exponential
distribution of pulses dominated by small pulses, thus larger
pseudo-residual distributions) to $\sim10^{-5}$ (flat distribution of
pulses, narrower pseudo-residual distributions), is found.



\section{Summary and Conclusions}

The unavoidable presence of residuals in the pulse shape prediction
used to compute OFC's is bound to introduce a systematic error in OF
amplitude estimates. This systematic error is independent of the pulse
amplitude and thus can be easily corrected, absorbing it in a simple
scale factor. Its size is maintained small (and even negligible) by
the usual requirements on the amplitude and shape of the pulse
prediction residuals. On the other hand, even the smallest residuals
will introduce a bias in the OF quality factor, impairing its
discriminating power.

The mathematical properties of the OF reconstruction do not allow the
measurement of the residuals shape and amplitude: this loss of
information is an intrinsic property of the OF reconstruction
technique itself. It is possible to extract from any data sample an
alternative vector of \emph{pseudo-residuals}, that allows the
correction of the original pulse shape prediction and obtain the exact
ionization pulse shape, apart from an uncertainty in the scale factor
that is bound to remain unknown. An ad hoc choice for the
normalization of the pseudo-residuals correction is shown to preserve
the original normalization of the reconstructed amplitudes. Given any
sample of signals from a readout cell it is always possible to obtain
its exact pulse shape from an initial (even imprecise) prediction,
while its absolute normalization must be obtained from different
sources. The pseudo-residuals correction restores the expected
$\chi^2$--like behavior of the OF quality factor and its full
discriminating power, independent of the initial bias.

The OF reconstruction properties discussed here, and the possibility
to obtain an optimal correction for any initial pulse shape
prediction, suggest a possible calibration strategy for the ATLAS EMC
at the LHC start-up: OFC's are computed exploiting the present best
knowledge of the ionization pulse shapes and normalizations, and as
soon as enough data is collected to compile reliable distributions of
the event pseudo-residuals, this technique is used to optimize OFC's
(thus quality factor distributions and noise reductions) without
spoiling the initial normalization of the reconstructed amplitudes.

\section{Acknowledgments}

The author is greatly indebted to Prof.~W.E.~Cleland, who first
suggested the possibility of computing a correction to the OFC's that
would preserve the initial normalization of the reconstructed
amplitudes, and then followed the development of this work with
continuous interest, encouragement, and very helpful discussions.

The author also wishes to thank R.~Froeschl for the useful discussions
and comments. The work presented here has been developed within the
ATLAS Liquid Argon Group, and the author thanks all the collaboration
members.



\bibliographystyle{elsart-num}

\clearpage

\begin{table}[t]
  \begin{center}
    \begin{tabular}{ccccccc}
      \hline \hline
      $i$ & $h_i$ & $g_i$ & $r_i$ & $R_{0i}$ & $a_i$ & $b_i$ \\
      \hline
      0  & 0.04761  & 0.04873 & -0.00111 &  1       & 0.14516 &  -7.2531 \\
      1  & 0.65923  & 0.67365 & -0.01442 &  0.07108 & 0.22650 & -26.9503 \\
      2  & 0.99769  & 0.99613 &  0.00156 & -0.15330 & 0.38105 &   9.1090 \\
      3  & 0.80987  & 0.80709 &  0.00279 & -0.29747 & 0.33019 &   5.9919 \\
      4  & 0.55451  & 0.55357 &  0.00095 & -0.10336 & 0.35092 &   8.3074 \\
      \hline \hline
    \end{tabular}
  \end{center}
  \caption{Numerical values used in the simulations.}
  \label{tab:PulseValues}
\end{table}


\end{document}